\documentclass[9pt,conference]{IEEEtran}
\IEEEoverridecommandlockouts

\usepackage{cite}
\usepackage{amsmath,amssymb,amsfonts}
\usepackage{algorithmic}
\usepackage{graphicx}
\usepackage{textcomp}
\usepackage{xcolor}
\usepackage{caption}
\usepackage{subcaption}
\usepackage{listings}
\usepackage{bbm}
\usepackage{graphicx}
\usepackage{multirow}
\usepackage{arydshln}
\usepackage{siunitx} 
\usepackage{url}
\usepackage{subcaption}
\usepackage{makecell}
\usepackage{bm}
\usepackage{mathtools}
\usepackage{float}
\usepackage[hidelinks]{hyperref}

\usepackage{array}    % Advanced table formatting
\usepackage{booktabs} % Better horizontal lines

\def\BibTeX{{\rm B\kern-.05em{\sc i\kern-.025em b}\kern-.08em
    T\kern-.1667em\lower.7ex\hbox{E}\kern-.125emX}}
\begin{document}

\title{Blind Spatial Impulse Response Generation from Separate Room- and Scene-Specific Information}

\author{\IEEEauthorblockN{Francesc Lluís}
\IEEEauthorblockA{\textit{Audio Technology Department} \\
\textit{Bang \& Olufsen A/S}\\
Struer, Denmark\\
frls@bang-olufsen.dk}
\and
\IEEEauthorblockN{Nils Meyer-Kahlen}
\IEEEauthorblockA{\textit{Dpt. of Information and Communications Engineering} \\
\textit{Aalto University}\\
Espoo, Finland\\
nils.meyer-kahlen@aalto.fi}
}

\maketitle
\begin{abstract}
For audio in augmented reality (AR), knowledge of the users’ real acoustic environment is crucial for rendering virtual sounds that seamlessly blend into the environment. As acoustic measurements are usually not feasible in practical AR applications, information about the room needs to be inferred from available sound sources. Then, additional sound sources can be rendered with the same room acoustic qualities. Crucially, these are placed at different positions than the sources available for estimation. Here, we propose to use an encoder network trained using a contrastive loss that maps input sounds to a low-dimensional feature space representing only room-specific information. Then, a diffusion-based spatial room impulse response generator is trained to take the latent space and generate a new response, given a new source-receiver position. We show how both room- and position-specific parameters are considered in the final output.
\end{abstract}

\begin{IEEEkeywords}
Room Acoustics, Blind Room Acoustic Identification, Contrastive Learning, Diffusion Models.
\end{IEEEkeywords}

\section{Introduction}

\label{sec:intro}

For augmented reality (AR) telepresence, the essential acoustic task is to render virtual sound sources that seamlessly integrate into the real acoustic scene. To do so, virtual sources must be rendered binaurally using head-tracked headphones, incorporating the user's room acoustic environment \cite{gari_room_2022, neidhardt_perceptual_2022}. As performing dedicated measurements of every user's acoustic environment would not be feasible, estimating the room acoustics must be done blindly based on the audio signals in the real scene. Thus, blind estimation of perceptually valid spatial room impulse responses (SRIRs), which contain the temporal and directional information for rendering a virtual source at a specific source and listener position in a given room, is of paramount interest.

\subsection{Background}
Traditional DSP approaches for blind room impulse response (RIR) estimation often use cross-relation methods employing multiple distributed microphones \cite{xu_least-squares_1995, mei_adaptive_2015, jo_robust_2021}, or adaptive filtering techniques like least-mean-squares or frequency domain least-squares \cite{schneider_generalized_2016} that identify the response between a close reference and a more reverberant target. This general idea has also been applied to SRIR estimation where a so-called \emph{pseudo-reference} is obtained by far-field beamforming \cite{meyer-kahlen_blind_2022}, dereverberation \cite{deppisch2024blind}, or a combination of both \cite{deppisch2024smart}; the signals from several spatially distributed microphones are used as targets.

In addition, deep learning methods have been proposed for blind RIR estimation. They generally feature an encoder that extracts room acoustic information from a real reverberant signal and a generator that uses this information to synthesize the corresponding RIR. An early method following this approach was introduced by Steinmetz et al. \cite{steinmetz_filtered_2021}, and recently several new models have been proposed \cite{ratnarajah_towards_2023, liao_blind_2023, lee_yet_2023}. In addition, \cite{lee_differentiable_2022} follows the same general approach but employs practically relevant reverberators, like feedback delay networks or filtered velvet noise at the generation stage.

Alternatively, algorithms referred to as blind parameter estimation do not aim at generating a complete RIR, but instead output well-interpretable room acoustic parameters such as reverberation time (RT) and direct-to-reverberant energy ratio (DRR). Many algorithms have been developed to address this task \cite{eaton_acoustic_2017, loellmann_comparative_2019}, including deep-learning-based approaches \cite{gamper_blind_2018, gotz_online_2023}. Given such parameter estimates, different generators may then be used to create a RIR, where the simplest, non-data-driven generator structure is exponential noise shaping\cite{moorer1979reverberation}. More advanced, data-driven generators include neural networks such as FAST-RIR \cite{ratnarajah_fast-rir_2022}. It remains unclear which parameters need to be estimated and what complexity of the generator is required to reach sufficient perceptual quality for a given application.

In summary, several approaches based on different encoders and generators for blind estimation of the full RIR have already been explored, either training them end-to-end or employing two separate algorithms for parameter estimation and RIR generation. 
However, existing approaches lack two features that are crucial to real-world AR applications.

Firstly, most approaches (including all deep-learning-based approaches introduced so far) have only been employed for single-channel RIR estimation; a RIR does not encompass the directional characteristics of acoustic environments inherent in the inter-channel relationships of multichannel SRIRs. Even though it is likely that not all spatial details of the response are relevant to achieving good perceptual quality, at least the correct direct sound direction and a plausible distribution of early reflections are required for binaural rendering.

Second, with the exception of \cite{lee_room_2023}, all methods aim at re-synthesizing one particular RIR based on one specific real sound source that needs to be active in isolation to achieve the most accurate estimation. In AR, however, users could easily find themselves in acoustic scenes containing multiple active sound sources. In addition, the goal is usually to render a sound source at new positions different from those occupied by real sources. Naturally, generating a physically accurate response for an unobserved position is unachievable, so the goal is instead to generate a new response that is perceptually plausible, ensuring that the rendered sound is perceived as if it is located at a new position in the same room. Therefore, it should encompass all room-specific and essential position-specific features.

\begin{figure}[ht]

\begin{minipage}[b]{1.0\linewidth}
  \centering
  \centerline{\includegraphics[width=\columnwidth]{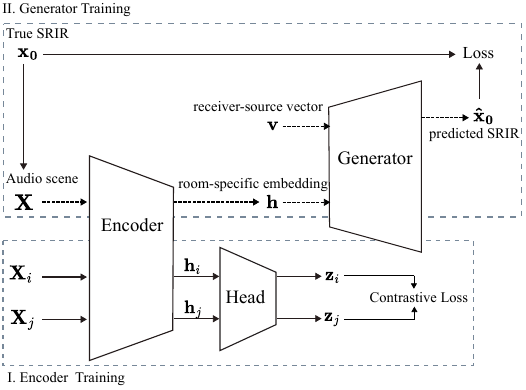}}
\end{minipage}
\caption{Training and inference stages. Initially, the encoder is trained using contrastive learning to capture all unique features of the room. During the generator's training, the encoder's weights are kept fixed. The generator, conditioned with a receiver-source position vector and a room-specific embedding, is then trained to produce spatial room impulse responses that incorporate both room-specific and position-dependent features. The inference path is indicated in dashed arrows.}
\label{fig:train_test_diagram}
\end{figure}

\subsection{Contributions}

In this paper, we present blind SRIR generation that is based on an encoder trained to encode only room-specific information, i.e., features that are invariant with respect to the specific location of the sources and receivers in the room. These room-specific features are extracted from entire acoustic scenes consisting of up to three sources rather than a single source. The encoder is trained using contrastive learning, similar to \cite{gotz_contrastive_2023}. Then, we show how the room-specific features at the encoder's output can be used together with scene-specific information, i.e., the positions of the source and receiver, to generate a SRIR using a diffusion-based generator. We work with four channel-SRIR that enable spatial analysis and rendering. We investigate three variants of the generator architecture, highlighting the most relevant choices. 
We analyze the generated SRIRs regarding reverberation time, resulting DRR, and Direction-of-Arrival (DoA) of the direct sound. Online listening examples highlight the perceptual feasibility of the approach\footnote{\url{http://research.spa.aalto.fi/publications/papers/icassp-srir-generation/}}.

\section{Methods}

\subsection{Dataset}
\label{subsec:data_gen}

We generate a synthetic dataset to train the model due to the lack of large, openly available datasets of SRIRs. However, we ensure that the generated data includes relevant variability. To make an informed choice of frequency-dependent RT, we analyze and extract octave-band RT profiles from the dataset of RIRs measured by Traer and McDermott~\cite{traer_statistics_2016}, that includes a natural distribution of relevant, everyday acoustic environments.

Next, we use pyroomacoustics \cite{scheibler_pyroomacoustics_2018} to simulate SRIRs. We randomly draw an RT profile from the RT distribution and select the wall materials that match the RT profile. Since the dataset in \cite{traer_statistics_2016} lacks information about room shapes, a common approach would be to assume shoebox-shaped rooms for simulation due to the simplicity of implementing image source algorithms as in \cite{gotz_contrastive_2023}. However, ideal, simulated shoebox-shaped rooms have unique features, such as audible ``sweeping echoes'' \cite{de_sena_modeling_2015}, that are uncharacteristic of real rooms and could introduce unwanted artifacts. To avoid this, we use slightly non-regular, hexahedron rooms in our dataset.

After drawing a random RT profile from the dataset, we randomly select the size of the room. The room dimensions are sampled uniformly, with the x and y dimensions ranging from 1.5 to 20~m, and the z dimension ranging from 2.5 to 8~m. We then approximate the required absorption coefficients using Sabine's equation.
If a specific RT profile/size combination is not realizable with absorption coefficients between 0 and 1, we draw a different room size.

Using this method, for each room we simulate two sets of SRIRs. Each set consists of one to three SRIRs truncated to 0.5 seconds at 48 kHz. We employ random sources positions and a single receiver position. We consider omnidirectional sources, while for the receiver we use a tetrahedral microphone array with a 2 cm radius and cardioid directivity patterns.
We then generate a pair of audio scenes for each room by convolving each SRIR with four-second-long signals from the AID \cite{gotz_aid_2022} database. In total, 45000 pairs of scenes are generated for training and 2500 for validation.

\subsection{Encoder}

The goal of the encoder is to predict an embedding that contains the room-specific information from a given audio scene. We train the encoder using a contrastive learning framework, where the neural network minimizes the distance between embeddings from scenes captured in the same room (positive examples) and maximizes the distance between embeddings from scenes captured in different rooms (negative examples).

As a preprocessing step, we encode the initial four seconds of the scenes using a short-time Fourier transform with a Hann window size of 512 and a hop size of 128. Next, we convert the complex valued time-frequency components into log-magnitude and instantaneous frequency components as in \cite{engel2019gansynth}, and apply standard normalization to each of the $4$ microphones. Finally, we collect the transformed time-frequency components of each microphone signal and flatten the data, yielding an input tensor \( \mathbf{X} \in \mathbb{R}^{(4 \cdot 2)  \times t \times f} \).

The encoder network consists of nine stacked residual convolutional blocks. Each block uses $3 \times 3 $ kernels and halves the input features using a stride of 2. Each residual convolutional block contains two convolutional layers with batch normalization and ReLU activation, along with a residual connection that includes a convolutional layer and batch normalization. The first convolutional block expands the 8 input data channels to 128 channels, while the subsequent blocks maintain 128 channels throughout. 

The convolutional blocks operate as a feature extractor that processes the input to produce a single-dimensional representation of the frequency information. To also summarize the temporal content into a single dimension, we apply a max-pooling layer on the temporal dimension after the last convolutional block. This results in a embedding $\mathbf{h} \in \mathbb{R}^{128}$. This embedding is processed by a projection head with dropout regularization and a multi-layer perceptron (MLP) containing a 256-unit hidden layer with ReLU activation and an output layer of 128 dimensions, producing a latent representation $\bm z \in \mathbb{R}^{128}$.

The encoder network is trained using the NT-Xent loss \cite{chen_simple_2020}
\begin{align}
    \mathcal{L} = -\log \frac{\exp(\bm z_i \cdot \bm z_j / \alpha)}{\sum_{k=1}^{2N} \mathbbm{1}_{[k \neq i]} \exp(\bm z_i \cdot \bm z_k / \alpha)},
\end{align}
where $i, j$ are the indices of a positive pair of examples, i.e., two scenes from the same room. $N$ corresponds to the batch size and $\alpha$ is a temperature scaling. In Fig.~\ref{fig:train_test_diagram}, this step is illustrated with the box labeled ``I. Encoder Training''. 

In the latent space constructed in this way, embeddings from the same room should be close together while embeddings from different rooms should be far apart. In addition, considering that the input scenes are generated using random source and receiver positions, the latent representation should capture only ``room-specific features'', independent of specific source and receiver positions.

The encoder network is trained for 125 epochs using the Adam optimizer. The learning rate is set to $3\times 10^{-4}$ and it is reduced by a factor 0.98 every two epochs. We use a batch size of 16 and the temperature scaling $\alpha$ is set to 0.1. After the training process, we select the weights with the lowest validation loss.

\subsection{Diffusion SRIR Generator}
\label{subsec:gen}

\begin{figure}[t]

\begin{minipage}[b]{1.0\linewidth}
  \centering
  \centerline{\includegraphics[width=\columnwidth]{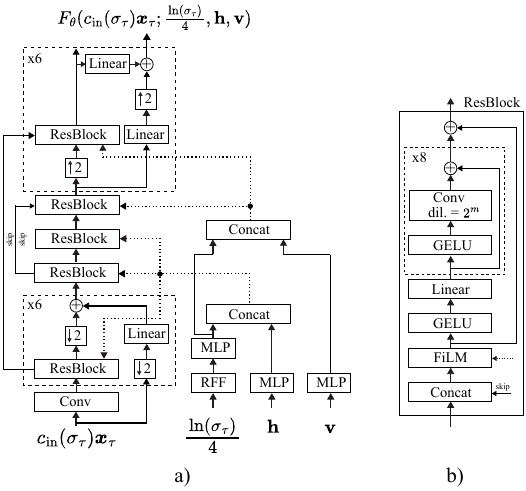}}
\end{minipage}
\caption{a) Diagram of the generator deep neural network architecture. b) Diagram of the Residual Block of the generator.}
\label{fig:gen_architecture}
\end{figure}

The goal of the generator is to generate a four-channel SRIR $\bm{\hat{x}_0}$  given the room-specific embedding $\mathbf{h}$ and the receiver-source vector $\mathbf{v} = \mathbf{s} - \mathbf{r}$, where $\mathbf{s}, \mathbf{r} \in \mathbb{R}^{3}$ are arbitrary source and receiver positions in Cartesian coordinates normalized by the largest room dimensions in the training dataset.

To this end, we propose a diffusion model \cite{NEURIPS2020_4c5bcfec, NEURIPS2019_3001ef25} with variance exploding parameterization and precondition strategy from Karras {\it et al.}~\cite{karras2022elucidating}. Diffusion methods model the data distribution by gradually adding noise to samples $\bm{x_0}$ from $p_\text{data}$ until they diffuse into Gaussian noise, and then learning to reverse this process to generate new data from noise. The generator network $D_\theta$ is trained using a denoising score matching objective, minimizing: 
\begin{equation}\label{eq:loss}
    \mathbb{E}_{\bm{x_0} \sim p_\text{data}, \boldsymbol\epsilon \sim \mathcal{N}(\mathbf{0},\mathbf{I}) }  \left[ \lambda(\sigma_\tau) \lVert D_\theta(\bm{x_0}+\sigma_\tau\bm{\epsilon};\sigma_\tau, \mathbf{h},\mathbf{v}) -\bm{x_0}   \rVert_2^2 \right],
\end{equation}
where $\tau$ indexes the diffusion steps, $\sigma_\tau$ is the noise standard deviation, and $\lambda(\sigma_\tau)$ is a weighting function. Following ~\cite{karras2022elucidating}, we further express the denoiser $D_\theta$ as 
\begin{multline}
D_\theta(\bm{x}_\tau; \sigma_\tau, \mathbf{h}, \mathbf{v})=c_\text{skip}(\sigma_\tau)\bm{x}_\tau+\\
c_\text{out}(\sigma_\tau)F_\theta(c_\text{in}(\sigma_\tau)\bm{x}_\tau; \tfrac{\ln(\sigma_\tau)}{4}, \mathbf{h}, \mathbf{v}),
\end{multline}
where $\bm{x}_\tau = \bm{x_0}+\sigma_\tau\bm{\epsilon}$ is the noise perturbed SRIR and $F_\theta$ is the neural network to be trained. $c_\text{skip}(\sigma_\tau)$, $c_\text{out}(\sigma_\tau)$, and $c_\text{in}(\sigma_\tau)$ are learned scaling parameters and the weighting function  $\lambda(\sigma_\tau)$ from eq. \eqref{eq:loss} is set to $1/c_\text{out}(\sigma_\tau)^2$. During training, we use ground truth SRIRs from the generated dataset, normalized to their maximum absolute value. In addition, SRIRs are aligned to start at $t=0$ to ensure the generator focuses solely on modeling the SRIR itself rather than the Time-of-Arrival (ToA), which can be easily inferred from the receiver-source vector $\mathbf{v}$. However, we also include a variant that is trained without alignment, predicting ToA as well.

\setlength{\tabcolsep}{8.5pt}
\begin{table*}[h]
\caption{Model performance on Mid-Frequency reverberation time (RT), broadband direct-to-reverberant energy ratio (DRR), and Direction-of-Arrival (DoA) of the direct sound. Results shown as root mean squared error (RMSE), correlation coefficient ($\rho$), bias, and percentage of data contained with the just noticeable difference (\% in JND).}
\centering
\begin{tabular}{lcccccccccc}
\toprule
\textbf{Model}                   & \multicolumn{4}{c}{\textbf{Mid-Frequency RT (s)}} & \multicolumn{4}{c}{\textbf{DRR (dB)}} & \textbf{DoA Error ($^\circ$)} \\ 
\cmidrule(lr){2-5} \cmidrule(lr){6-9}
& \textbf{RMSE}  & \textbf{$\rho$}  & \textbf{BIAS} & \textbf{$\%$ in JND} & \textbf{RMSE}  & \textbf{$\rho$}  & \textbf{BIAS} &  \textbf{$\%$ in JND} &  \\ \midrule
PROPOSED                      & 0.206 & 0.738 & 0.049 & 27.8      &  2.62 & 0.872 & 1.10 & 83.3 & 3.46
\\
\textit{variant}: CONCAT ALL EMB            & 0.212 & 0.711 & 0.028 & 25.4                &2.99 & 0.860 & 1.61 & 77.5 & 3.46 \\
\textit{variant}: WITH ToA                         & 0.204 & 0.734 & 0.030 & 28.7              & 3.10 & 0.869 & 2.00 & 78.9 & 4.58\\ \bottomrule
\end{tabular}
\label{tab:merged}
\end{table*}

To sample a SRIR, we use the stochastic sampler proposed by \cite{karras2022elucidating} where the noise level is defined as $\sigma_\tau = \tau$ and distributed according to
\begin{equation}\label{schedule}
    \tau_{i<T}=\left(\sigma_{\text{max}}^{\;\;\;\frac{1}{\rho}} 
    + \tfrac{i}{T-1}\left(
    \sigma_\text{min}^{\;\;\;\frac{1}{\rho}}
    -\sigma_\text{max}^{\;\;\;\frac{1}{\rho}}
    \right)\right)^\rho,
\end{equation}
where $T$ represents the total number of discretized steps and $i$ indicates the discretization index. We use $\rho=10$, $\sigma_\text{min}=10^{-6}$, $\sigma_\text{max}=10$, and $T = 35$. We set the stochasticity parameter of the sampler as $S_{\text{churn}}=1$.

The Generator architecture is shown in Fig.~\ref{fig:gen_architecture}. The architecture is inspired by \cite{moliner2023solving} and follows a U-Net-like deep neural network, i.e. an encoder-decoder architecture with skip connections. Each encoding block applies a Residual block consisting of an initial FiLM layer, a stack of exponentially-increasing dilated convolutional layers with GeLU activations and residual connections. Then, the time features are downsampled by a factor of 2. Each decoder block first upsamples the features and then applies a Residual block. The bottleneck consists of a residual block with no upsampling/downsampling.

FiLM conditioning in the encoder blocks and the bottleneck is driven by the noise level and $\mathbf{h}$, while in the decoder blocks is driven by the noise level and $\mathbf{v}$. These factors, with the noise level first encoded using Random Fourier Features as in \cite{moliner2023solving}, are processed by an MLP and concatenated. The MLPs employ two hidden layers of 128, 256 units, ReLU activations and a 512-dimension final layer. We also experimented with concatenating the three embeddings to drive FiLM layers.

The generator network is trained for 300 epochs using the Adam optimizer with an initial learning rate of $3\times 10^{-4}$, reduced by a factor 0.8 every ten epochs. We use a batch size of 8 and select the weights with the lowest validation loss.

\begin{figure*}
    \centering
\includegraphics[width=\textwidth]{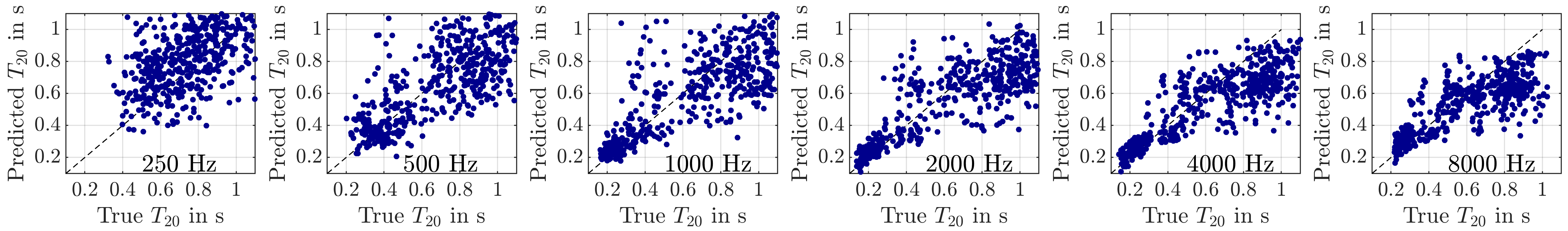}
    \caption{Reverberation time in octave bands determined from the generated and the true responses for the PROPOSED model. Underestimation occurs mainly at high RTs. \vspace{-0.5cm}}
    \label{fig:rts}
\end{figure*}

\begin{figure}[t]
    \centering
\includegraphics[width=\columnwidth]{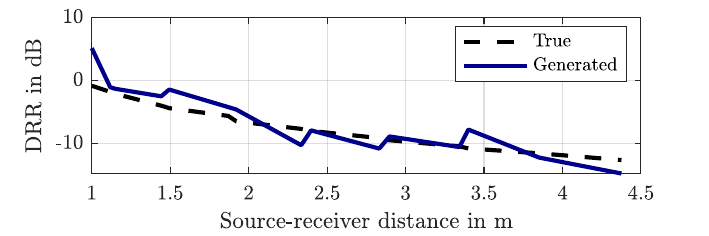}
    \caption{Example of true DRR and DRR of the generated SRIR along a line leading past the source for the PROPOSED model.}
    \label{fig:drr-example}
\end{figure}

\begin{figure}[t]
\vspace{-0.5cm}
    \centering
    \includegraphics[width=\columnwidth]{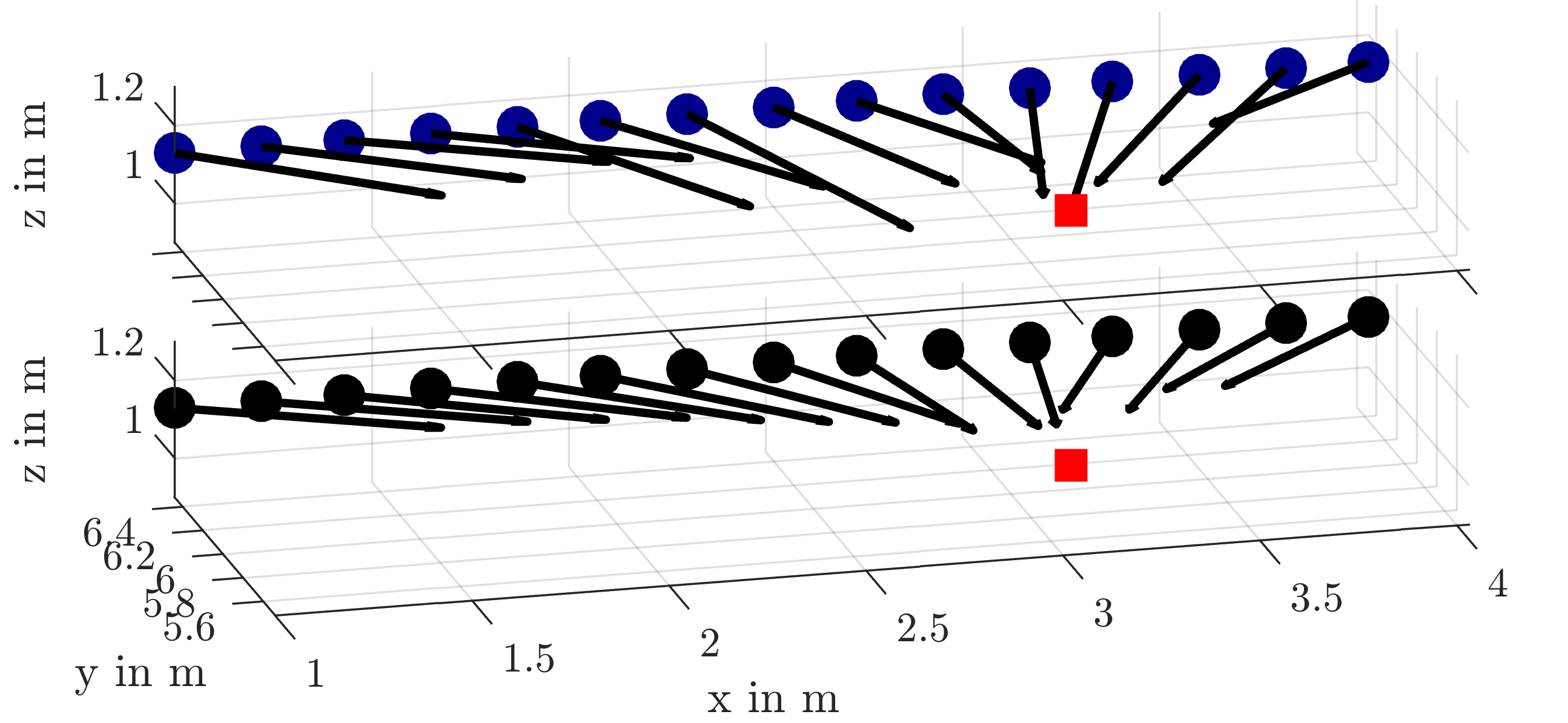}
    \caption{Example of direct sound DoAs represented by arrows. Receiver positions as dots, source position as red square. The DoA obtained from generated responses on above and ground truth DoAs below. Results are shown for the PROPOSED model.}
    \label{fig:doa-example}
\end{figure}

\section{Results}

We analyze the performance regarding acoustic parameters that are either expected to be room-specific or position-specific. In each case, we show the performance of our PROPOSED generator (explained in Sec. \ref{subsec:gen}), but also indicate the performance of the generator when concatenating the three embeddings to drive FiLM layers in all blocks of the network (CONCAT ALL EMB). Also, we forgo time alignment during training for one of the variants, so the generator needs to consider ToA as well (WITH ToA).

The performance is evaluated on a test set generated following a similar procedure as the training data (see Sec. \ref{subsec:data_gen}). However, a different set of RT profiles is used, obtained from a set of 80 rooms measured at Aalto University. In total, the test set is composed of audio scenes from 100 different rooms. Additionally, we simulate the ground truth SRIR for these rooms at 15 receiver positions along a line near a source position.

\subsection{Room-specific parameter: Reverberation Time}

In simple rooms like the ones used here, the reverberation time can be considered independent of the specific source and receiver positions.
Fig.~\ref{fig:rts} shows reverberation time estimates of the ground truth and the predicted responses for octave bands, computed in accordance with ISO 3382 \cite{iso_338212009_acousticsmeasurement_2009}. Results in Tab.\ref{tab:merged} correspond to the mean value between 500~Hz and 1~kHz bands, as it is often used as a single value in room acoustics. The highest correlation coefficient is $\rho = 0.738$, which is obtained for the PROPOSED generator. Determining how many responses are within a relatively tight $10\%$ margin of assumed just noticeable difference (JND), as done in \cite{lee_room_2023}, around the true RT leads to results between $25.4\%$ and $28.7\%$.

These results indicate that the encoder trained using the contrastive loss can extract room-specific information from complete acoustic scenes and that the generator can be conditioned upon this information.

\subsection{Position-specific parameters}

Now that we have shown that the generated SRIR captures the frequency-dependent reverberation time, we assess whether the generator's conditioning incorporates the position-specific information in the generated responses as well. For this, we run inference for 15 receiver positions on a line passing by the source at a distance of 1~m. Based on this data, we analyze two position-dependent parameters.

First, we assess the DRR, which strongly depends on the distance between the source and the receiver. It is computed by taking the ratio of the energy contained in the 1.25~ms around the first peak and the rest of the response. 
One example is shown in Fig.~\ref{fig:drr-example}. Therein, it becomes clear that the DRR of the responses generated along the line approximates the actual DRR decay. Tab.~\ref{tab:merged} shows the DRR errors evaluated at 15 positions for each of the rooms in the test set. For the PROPOSED model, $83.3\%$ of responses are within JND. The JND for DRR depends on the true DRR; in \cite{larsen_minimum_2008} it was found to be 2.4~dB when the true DRR was 0~dB, with an increase to both lower and higher true DRR. Piece-wise linear approximation of the results shown in \cite{larsen_minimum_2008} is used as a threshold function here.

Second, we check whether the DoA of the direct sound points to the source position that was provided during inference. The DoA is estimated using the Time Difference of Arrival between the microphone responses, as in \cite{tervo_spatial_2013}. Fig.~\ref{fig:doa-example} shows the DoA of the direct sound observed at each of the 15 receiver positions. For the PROPOSED model, they point towards the source within a mean great circle distance of 3.46$^\circ$, see last column in Tab.~\ref{tab:merged}. Note that these DoA results indicate not only that position-specific information was learned, but also that the inter-channel relations of the generated SRIR give valid directional estimates in the first place.

\section{Conclusion and Future Work}

We proposed a method for blind SRIR generation that first extracts room-specific features from a complete acoustic scene and then generates SRIRs at unobserved positions given arbitrary source and receiver positions. While these generated SRIRs cannot be physically accurate, objective metrics indicate that they capture both room and position-specific features.

With the AR application in mind, a listening experiment needs to be conducted as a next step; it should determine whether listeners are unable to detect if a sound source is rendered using a generated response when compared to other sources rendered at different positions using ground truth responses. In the meantime, such examples are provided online.

Finally, future versions of the model should be trained and tested using measured data, for which new, large-scale SRIR datasets would be required.

\end{document}